\documentclass[sensors,article,submit,pdftex,moreauthors]{Definitions/mdpi} 

\firstpage{1}  
 \pubvolume{1} \issuenum{1} \articlenumber{0}
\pubyear{2024} \copyrightyear{2024}
\datereceived{ } 
\daterevised{ } % Comment out if no revised date
\dateaccepted{ } 
\datepublished{ } 
\hreflink{https://doi.org/} % If needed use \linebreak
\preto{\abstractkeywords}{\nolinenumbers}
\Title{Smartphone
  light sensors as innovative tool for solar irradiance measurements}

\TitleCitation{Smartphone light sensors as innovative tool}

\Author{José Luis Di Laccio$^{1}$\orcidA{},
          Andrés Monetta$^{1}$\orcidB{},
          Rodrigo Alonso-Suáreza$^{2}$\orcidC{},
          Martin Monteiro$^{3}$\orcidD{},
          and Arturo C. Marti$^{4}$\orcidE{}}

\AuthorNames{José Luis Di Laccio, Andrés Monetta, Rodrigo Alonso-Suárez, Martin Monteiro,
 and Arturo C. Marti}
 
\AuthorCitation{Di Laccio, J.L.: Monetta, A.; Alonso-Suarez, R.; Monteiro, M.; Marti, A.C.}

\address{%
$^{1}$ \quad  Departamento de
  Física, CENUR Litoral Norte, Universidad de la República, Salto,
  Uruguay\\
$^{2}$ \quad  Facultad de Ingenieria,
  Universidad de la República, Montevideo, Uruguay\\
$^{3}$ \quad  Facultad de Ingenieria,
  Universidad ORT Uruguay, Montevideo, Uruguay\\
$^{4}$\quad  Facultad de Ciencias,
  Universidad de la República, Montevideo, Uruguay\\
  }

\corres{Correspondence: marti@fisica.edu.uy.}

\abstract{The use of solar energy has experienced explosive growth in
  recent years due to the comparative advantages it offers.  Here, we
  present an innovative tool for using the light sensor of smartphones
  for solar irradiance experiments in particular to measure the
  primary component of ground-level solar radiation: the direct normal
  solar irradiance. We provide comprehensive guidelines for using this
  sensor in a variety of situations and show how to calibrate it using
  easy available clear-sky satellite data. In addition, as an
  alternative approach, we also incorporate a comparison with
  professional reference measurement estimates to demonstrate the
  consistency of the proposed methods. This experiment can be
  performed with the vast majority of available smartphones, some
  readily available materials plus an open location in clear sky
  conditions for a lapse of a couple of hours.  In this way, the
  smartphone becomes a valuable tool for measuring the solar resource
  and to propose several experiments focusing on the physics of solar
  radiation and its interaction with the Earth's atmosphere, exploring
  solar variations across locations, cloud conditions, and time
  scales.}


\begin{document}

\section{Introduction}
Photovoltaic energy has experienced spectacular growth in recent years
\cite{BP-2021,REN21-2021}. This growth contrasts with the little
attention given to solar resources in science and engineering
curricula.  In general, the study of solar radiation is predominantly
approached from a theoretical standpoint, potentially due to the
technical complexities and high costs associated with measurement
instruments. As a result, experiments involving the measurement of
solar irradiance and the determination of its uncertainty are
typically reserved for advanced programs or specialized
laboratories. It is also worth mentioning that this subject matter
provides a platform to develop various cross-cutting competencies,
including measurement techniques and environmental stewardship.

The main objective of this work is to show how to calibrate a light
sensor like the ones incorporated in most smartphones to measure the
solar resource.  The achievement of this objective allows to dispose
of an instrument to measure solar radiation for other activities
and/or laboratories and also provides the opportunity to learn about
the solar resource in general. 
In particular, we show how to measure broadband direct solar
irradiance at normal incidence (DNI), the main component of solar
radiation at ground level. Direct irradiance is the portion of the
incident radiation that arrives directly from the solar disk without
being absorbed or scattered in the Earth's atmosphere. This component
is essential to evaluate the performance of concentrating solar
applications and to estimate the solar irradiance available on an
inclined plane, necessary for solar photovoltaic and low-temperature
solar thermal applications in which it is usual practice to tilt the
solar collection surfaces.

The use of smartphones as a measurement
instrument, which have also shown innovative contributions to the
teaching of Physics due to the availability of various built-in 
sensors (for  recent reviews see Ref.~\cite{monteiro2022resource,GilDila}).
Although the type of sensor varies 
the type of sensor varies considerably depending on the specific model, in general, acceleration, magnetic field or luminosity sensors are the most common. 
The latter, which we will use in this work, has been used in relatively few experiments. It is worth mentioning its use for experimentation with polarized waves \cite{monteiro2017polarization}
or in problems motivated in astronomical situations \cite{DiLaccio2021}. Among the advantages of using smartphone sensors we can mention that it is possible to measure simultaneously with several sensors \cite{monteiro2019physics}
or supplementted with video analysis \cite{monteiro2021smartphone}. In the framework of experiences aimed at students, the determination of uncertainty and the study of the fluctuations of these sensors cannot be left aside \cite{Monteiro2021Using}

To reach our goal, we describe two procedures for calibrating the
sensor, one based on a professional reference measurement and the
other using publicly-available satellite estimates of ground level
clear-sky irradiance. We compare the calibration procedures and show
that both are feasible. By bridging this gap about calibration, we
enable the measurements and instructional laboratory to be done with
only the smartphone and manual positioning, if required.  We must
emphasize that the professional equipment mentioned in the paper,
including tracking, is not indispensable to carry out the proposal,
and is included only to show the reader that the smartphone is a
calibrable and reliable instrument.

As the topics concernig the solar resource are not usually covered in
physics courses, next section introduces the various magnitudes and
usual geometric calculations in the area of solar energy and
lighting. Section~\ref{materiales_metodos} shows the experimental
setup carried out in this work for the calibration of the illuminance
sensor of a smartphone for the measurement of DNI. This section also
introduces the solar satellite estimates, which are used here as an
alternative reference data set for smartphone calibration. Section
\ref{calibracion} describes the calibration process and its
uncertainty evaluation.  Finally, Section \ref{conclusion} summarizes
the conclusions.

\section{Theoretical framework: solar radiation}\label{teorico}

\subsection{Basis of the interaction between solar radiation and the atmosphere}
The solar irradiance, $G$, is the incident power per unit normal
surface whose beam comes from the Sun. The solar irradiance at the top
of the atmosphere, $G_{o}$, varies because of two factors: a variation
of about $\pm3$\% due to the elliptical character of the Earth's orbit
around the Sun, and also, it present small variations due to
oscillations in the solar activity, typically below $0.3$\%
\cite{Frohlich-2006}. The averaged solar irradiance at the top of the
atmosphere on a surface normal to the Earth-Sun direction and when the
Earth is at a distance equal to the mean Earth-Sun distance (1
Astronomical Unit or AU) is known as the solar constant \cite{Kopp11},
$G_{sc}=1361$~Wm$^{-2}$. In this way, the seasonal variation of $G_o$
is obtained by multiplying the solar constant by the orbital factor,
$F_{n}=r^{-2}$, accounting for the variation of the Earth-Sun distance
and it can be approximated with an uncertainty of 0.25\% by
\begin{equation}\label{ec1}
F_n = 1 + 0.033 \times \cos\left( \frac{2\pi n}{365} \right), 
\end{equation}
where $n$ is the ordinal day number (going from 1, Jan. 1st. to 365
Dec. 31) \cite{Iqbal,Duffie2013}. The extraterrestrial irradiance at
normal incidence is obtained then as $G_o = G_{sc} \times F_n$.

Once solar irradiance penetrates the Earth's atmosphere, it interacts
with various atmospheric components such as air, aerosols, water
vapor, and cloudiness. This interaction leads to scattering in
multiple directions, with some of the irradiance being absorbed by
these components, while the remaining portion is reflected back into
space.  The combination of these two components on a horizontal plane
is known as the global horizontal irradiance (GHI), denoted as
$G_{h}$, which represents the solar energy magnitude most commonly
measured on the Earth's surface. Several methods can be employed to
measure this quantity, including the use of photodiodes, calibrated
photovoltaic cells, or thermopile pyranometers. Among these options,
the last offers the highest precision.

Direct normal irradiance, denoted as $G_{bn}$, is less frequently
measured since its continuous measurement requires fine solar tracking
mechanisms that ensure the measuring equipment to be aligned at all
times pointing to the solar disk. The measuring instrument, the
pyroheliometer, is equipped with a collimating tube that filters any
irradiance that does not come from its normal direction, with a
convention aperture of 5~sr of solid angle, corresponding to a typical
solar disk. The size of the solar disk observed from Earth depends on
atmospheric conditions. In the presence of high humidity, for example,
the perceived solar disk enlarges due to a larger size of the
circumsolar region. The solid angle of 5~sr associated with the solar
disk is, in effect, a convention value. The standard that classifies
solar radiation measurement instruments is the ISO 9060:2018 which
establishes categories according to the quality of the equipment
(offset, angular error, response time, among others) and the
corresponding uncertainty. After measuring the DNI, the atmospheric
transmittance can be estimated as $T_{b} = G_{bn} / G_{o}$, a quantity
that enables understanding the amount of available solar radiation for
capture and conversion into solar energy.

\begin{figure*}[htbp!]
\centering
\includegraphics[width=0.45\textwidth]{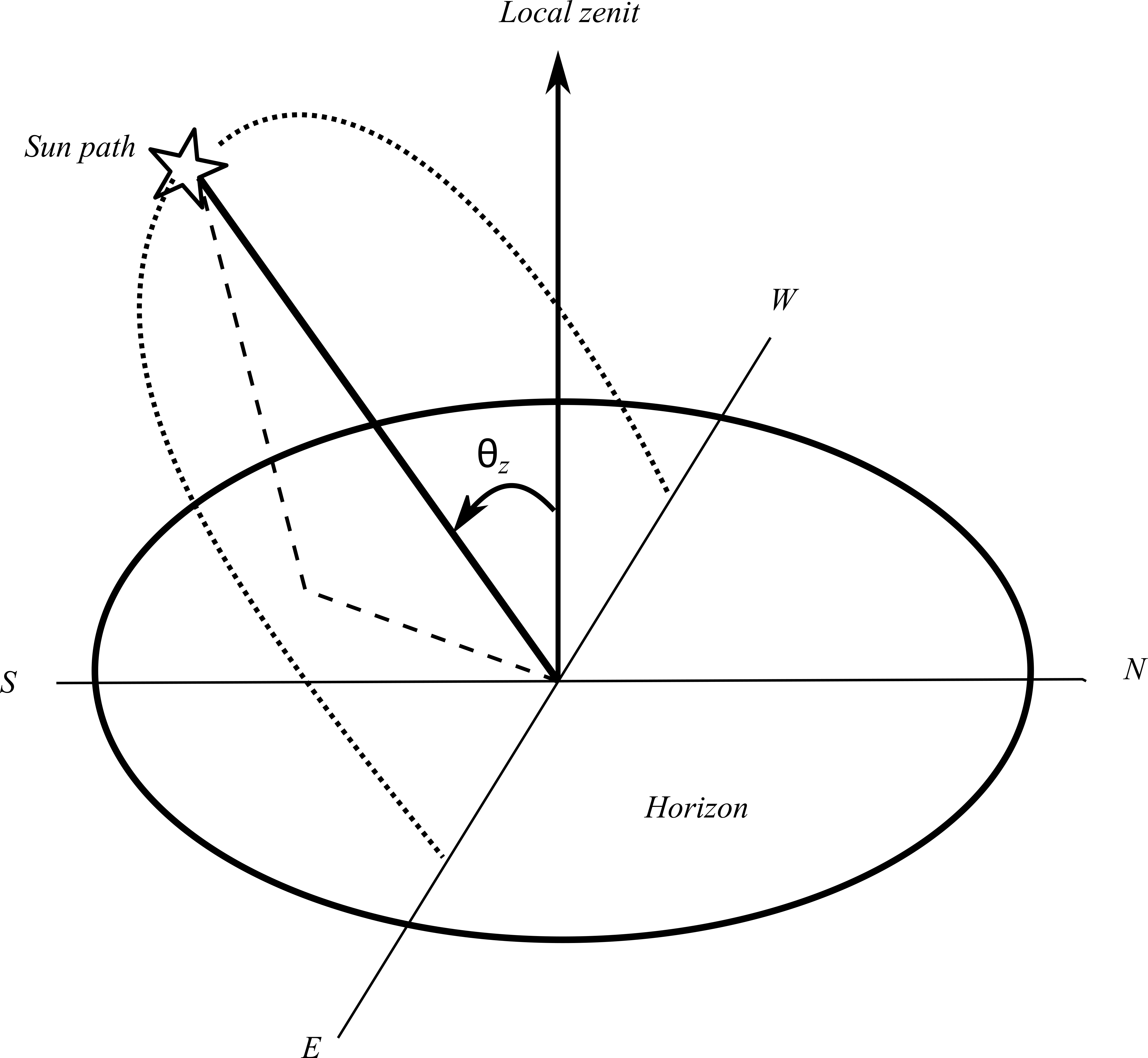}
\includegraphics[width=0.48\textwidth]{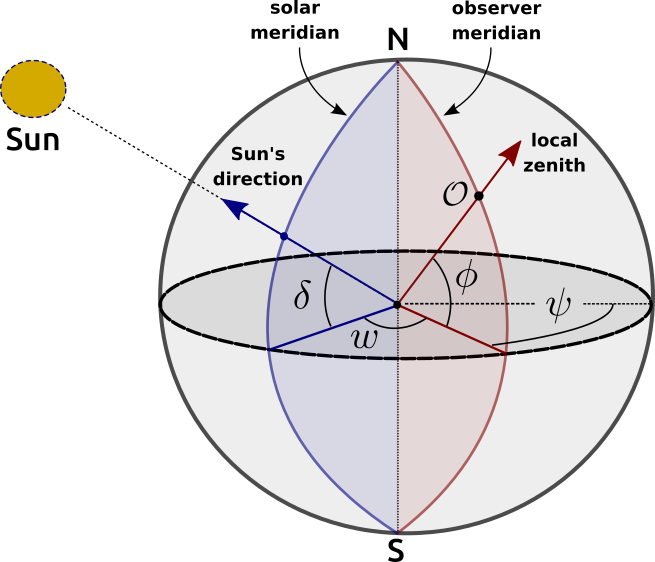}
\caption{Schemes of the different angles related to the Sun's apparent
  movement. ; the solar zenith angle $\theta_z$ (left) and $\delta$,
  $\phi$, $\psi$ (right).  }
\label{fig:diagramas}
\end{figure*}

It is useful to define the main angles related to solar radiation.
One of them is the solar zenith angle, denoted as $\theta_z$, which
represents the angle between the direction of the Sun and the local
vertical (referred to as the local zenith) as shown in
Fig. \ref{fig:diagramas}. The cosine of this angle appears recurrently
in expressions related to solar radiation, especially for magnitudes
projected onto the horizontal plane, and its calculation is carried
out according to \cite{Duffie2013}
\begin{equation}
\cos\theta_z = \cos\phi \cos\delta \cos\omega + \sin\phi \sin\delta,
\label{ec3}
\end{equation}
where $\phi$ is the latitude, $\delta$ is the solar declination angle
and $\omega$ is the hour angle shown in
Fig. \ref{fig:diagramas}. Latitude is the angle between the Earth's
equator (parallel 0°) and the site of interest (indicated by
$\mathcal{O}$ in the figure), along the observer's meridian. By
convention, latitudes are positive north of the equator and negative
south of the equator. Solar declination is the angle formed by the
Earth-Sun line with the Earth's equatorial plane, and can be
calculated in radians with good precision through the expression:
\begin{equation}
\delta= 0.4095\times\sin\left( 2\pi~\frac{(n+284)}{365}\right).
\label{ec4}
\end{equation}
Finally, the hour angle is the angle on the equatorial plane between
the observer's meridian and the solar meridian. This angle varies with
the apparent position of the Sun respect to the Earth and it is
calculated from the time label associated with each
measurement. Figure \ref{fig:diagramas} also depicts a fourth relevant angle,
$\psi$, the longitude of the observer measured from the Greenwich
meridian (defined as $\psi=0$°).

The hour angle is related to the solar time of the site, $t_{s}$, according to
\begin{equation}
\omega = \frac{\pi~(t_{s}-12)}{12}. 
\label{ec5}
\end{equation}
Indeed, this angle vanishes at solar noon ($t_{s} = 12$~hours),
i.e. when the solar meridian coincides with the observer's meridian,
and grows at a rate of $\pi/12$ radians per hour, the speed of
rotation of the Earth. To complete the calculation, it remains to link
the solar local time with the local standard time, $t_u$, expressed
according to a given UTC time zone associated to a central meridian,
$\psi_{u}$. For example, the time in UTC-3 is associated with a
meridian of $\psi_{u} = -45$°. The relationship between both hours
includes the so-called \textit{equation of time}, $E$, and is defined
by
\begin{equation}
t_{s} = t_{u} + \frac{E + 4~(\psi-\psi_{u})}{60}, 
\label{ec6}
\end{equation}
where $t_{u}$ is the local standard time expressed in hours and
fractions, $\psi$ and $\psi_{u}$ are the signed longitude in decimal
degrees (negative for West longitudes and positive for East
longitudes) of the site and the reference UTC, respectively, and $E$
is expressed in minutes. $E$ can be calculated as a Fourier expansion
as proposed by Spencer \cite{Spencer-1971}.

\subsection{Attenuation in the atmosphere}

The mass of air or relative optical mass, $m_{i}$, is a dimensionless
quantity that is defined as the quotient between the amount of mass of
a certain component $i$-th of the atmosphere that a beam of radiation
covers in its trajectory and the one that it would cover in a vertical
path, that is, in the direction of the zenith.  Under the hypothesis
of a non-anisotropic flat atmosphere, we can assume $m_{i} = m
=1/\cos\theta_z$ \cite{Iqbal,Duffie2013}. The uncertainty associated
with this expression due to neglecting the terrestrial curvature and
the refraction phenomena grows as the zenith angle is larger. However,
the expression presents an uncertainty of about $0.25$\% for
$\theta_{z} = 60$° \cite{Iqbal}, and it is adequate for zenith angles
between 0° and 70°. There were proposed more precise expressions that
can be used for large zenith angles, around 80-90°, such as that of
Kasten and Young \cite{KasterYoung1989}. Here, as we do not consider
measurements very early in the morning or very late at sunset the
previous expression, the previous approximation results appropriate.

The Lambert-Beer-Bourger law describes the attenuation of a direct
beam of radiation when passing through a medium. Its application to
the direct normal irradiance in the atmosphere results in an
exponential and spectrally selective attenuation
\begin{equation}
G_{b, \lambda} = G_{o, \lambda} e^{-m \tau_{\lambda}},
\label{eq:LBB_eq}
\end{equation}
where $G_{b, \lambda}$ is the direct spectral irradiance, $G_{o,
  \lambda}$ is the spectral extraterrestrial irradiance corrected by
the orbital factor, $\tau_{\lambda}$ is the optical depth of the
atmosphere and $m$ is the air mass. This equation can be derived from
the differential version of the Lambert-Beer-Bouger modeling the
atmosphere by a set of layers $i$, so as the transmissivity can be
expressed as $T_{i, \lambda} = \exp(-\tau_{i,\lambda }~m_i)$. The
total transmissivity results from the product of the layers, and
therefore $\tau_{\lambda}$ includes the effect of all different
components. This is a regular assumption when modeling the interaction
between the Sun's radiation and the atmosphere \cite{Gueymard-2008,
  Oumbe-2014}.

Clear-sky models adopt the Lambert-Beer-Bouger law to describe the
direct normal irradiance under ideal atmospheric conditions. In these
conditions, the attenuating components encompass various factors such
as air molecules (O$_2$, N$_2$, Ar), which are responsible for
Rayleigh scattering, as well as water vapor, aerosols, ozone, and
other minor gases. Ozone, although crucial for life on Earth due to
its role in attenuating ultraviolet radiation, has a relatively minor
contribution across the entire solar spectrum. A Rayleigh atmosphere
refers to a pristine and dry atmospheric state where only the
attenuation mechanism of molecular scattering is at play, leading to a
clean and transparent atmosphere.

Numerous clear-sky models have been developed based on these concepts
\cite{GUEYMARD20122145}.  Among them, the ESRA (European Solar
Radiation Atlas) model \cite{RIGOLLIER200033} strikes a favorable
balance between simplicity and precision, making it suitable for
implementation within the framework of an university experimental
laboratory. This model operates using the concept of global optical
depth, denoted as $\tau$, which encompasses the entire solar spectrum.
By incorporating the global optical depth of Rayleigh extinction,
denoted as $\tau_{R}$, we can express $\tau = \tau_{R} T_{L}$, where
$T_{L}$ represents Linke's turbidity which quantifies the number of
clean and dry atmospheres that would need to be stacked to achieve the
level of attenuation observed in the real atmosphere.  Consequently,
by adjusting a single parameter, $T_L$, based on ground measurements
it becomes feasible to construct a simple model for estimating DNI
under clear-sky conditions as $G_{bn} = G_{o} e^{-\tau_{R} m
  T_L}$. Among the several ways to approximate the Rayleigh optical
depth \cite{Kasten-1984, Louche-1986, KasterYoung1989, Grenier-1994,
  Kasten-1996}, here we use the Kasten's formulation
\cite{Kasten-1996}.
%, used in the ESRA model:
%\begin{equation}
%\frac{1}{\tau_{R}} = \left\{ \begin{array}{ll}
%6.62960 + 1.75130 m -  0.12020 m^2 + 0.00650 m^3 - 0.00013 m^4 \hspace{5mm}& m \leq 20 \\
%10.6 + 0.718 m & m > 20.
%\end{array}\right.
%\label{eq:tau_R}
%\end{equation}

\subsection{Illuminance}

Photometry is the area of knowledge that is responsible for measuring
the light perceived by the human eye. This quantity depends on the
sensitivity of the human eye to different wavelengths in the visible
region of the electromagnetic spectrum. Each wavelength has its
relative weight in the response of the human eye depending on the
lighting conditions (good or poor) in which the observer is. In
typical lighting conditions, corresponding to a real situation at Sun,
it is possible to relate the illuminance, $E_{v}$, measured in lumens
per unit area, lm/m$^2$ or lx, with the spectral irradiance
$G_{\lambda}$ according to
\begin{equation}\label{ec-E}
\displaystyle E_{v} = K_{m} \times \int_{380}^{780} G_{\lambda}~V_{\lambda}~\mathrm{d}\lambda, 
\end{equation}
where $V_{\lambda}$ is the spectral response of the human eye and
$K_{m} = 683~\mathrm{lm/W}$ is the maximum luminous efficacy obtained
with monochromatic illumination at $\lambda = 555 $~nm.

Consequently, establishing a precise relationship between illuminance,
denoted as $E_v$, and broadband solar irradiance is not a
straightforward task, as it depends on the spectral composition of
solar irradiance at ground level within a specific portion of the
spectrum. This composition, in turn, is influenced by the atmospheric
conditions. This scenario resembles the calibration process for
photovoltaic radiometers used to measure solar irradiance. These
devices possess distinct spectral responses across different regions
of the solar spectrum and are calibrated by comparing them to
pyranometric radiometers with a flat spectral response (broadband)
encompassing the entire solar spectrum. As a first approximation,
these spectral differences can be disregarded, and the customary
approach involves employing a constant or global calibration curve,
determined under clear-sky conditions, to account for these effects
\cite{ISO9847}. This calibration methodology is adopted in the present
study.

\section{Materials and methods}\label{materiales_metodos}
The objective of this work is to demonstrate the usefulness of
smartphones as an experimental measurement tool for direct normal
solar irradiance. This requires mounting a tube around the smartphone
light sensor, pointing it directly at the Sun, and then calibrating
its measurement.
%After calibration, the smartphone can be used to
%determine parameters of the state of the atmosphere, such as direct
%transmittance, $T_b$, or Linke's turbidity, $T_L$.
In this study, the calibration of the equipment is conducted using two
distinct approaches: (i) by comparing it to high-quality pyrheliometer
data obtained from professional measurements, and (ii) by comparing it
to estimates from sophisticated publicly available clear-sky
models. Both calibration methods require clear-sky conditions to
ensure consistent measurements and to mitigate any discrepancies
associated with cloud movement.

Calibration method (i) demonstrates the potential of using smartphones
for direct DNI measurements, as it utilizes a reference instrument of
Secondary Standard quality. This reference instrument exhibits a
measurement uncertainty of less than 1\%, and it is calibrated with
traceability to the World Primary Standard (WSG) at the World
Radiation Center (WRC) in Davos, Switzerland. This calibration
approach validates the use of smartphones as measurement devices for
DNI, providing a robust and reliable reference for comparison.

Calibration method (ii) offers an alternative approach for calibrating
smartphones in situations where terrestrial reference measurements are
unavailable. This alternative method allows for the widespread use of
smartphones as measurement instruments on a large scale and at a low
cost. It utilizes sophisticated clear-sky models, which serve as a
general calibration reference for smartphones. This approach addresses
the need for smartphone-based measurements when traditional
terrestrial reference measurements are not feasible.

\subsection{Experimental measurements}

\begin{figure*}[ht!]
\centering
\includegraphics[height=0.37\textwidth]{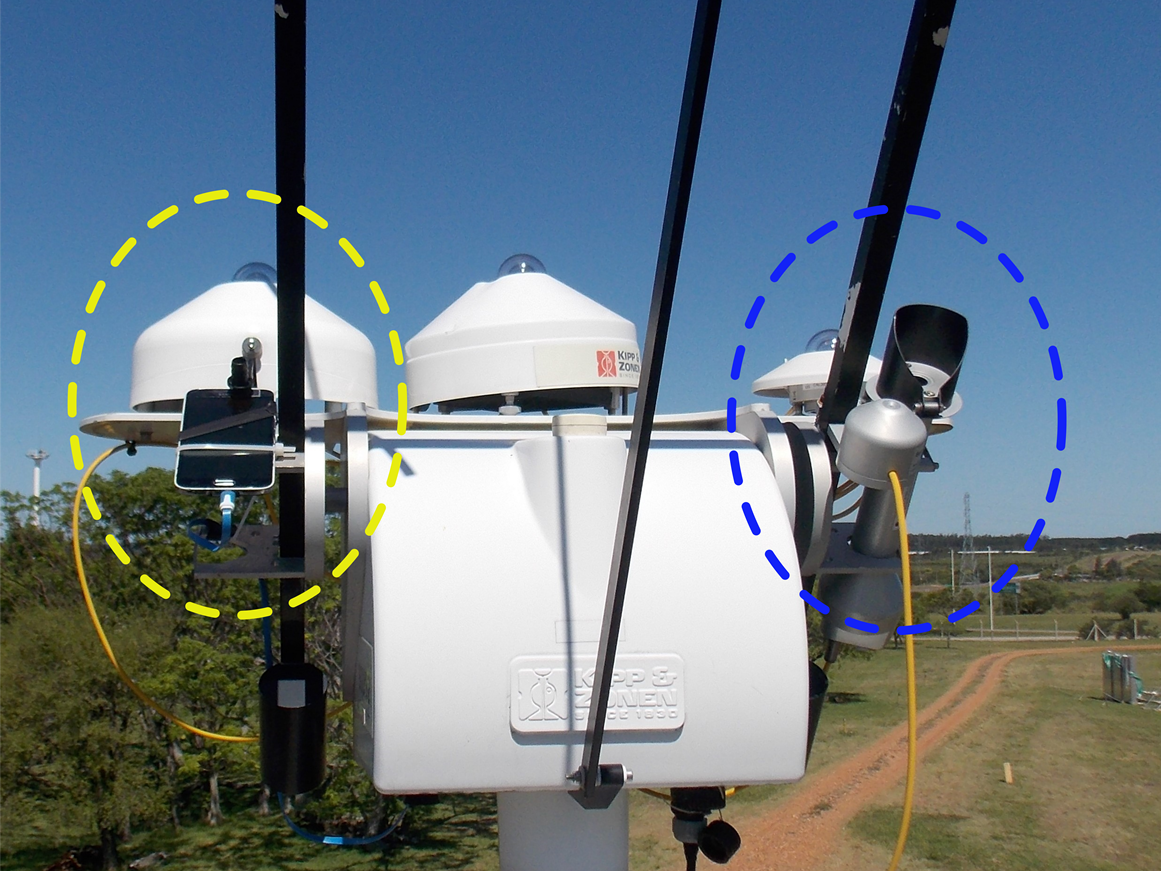}
\includegraphics[height=0.37\textwidth]{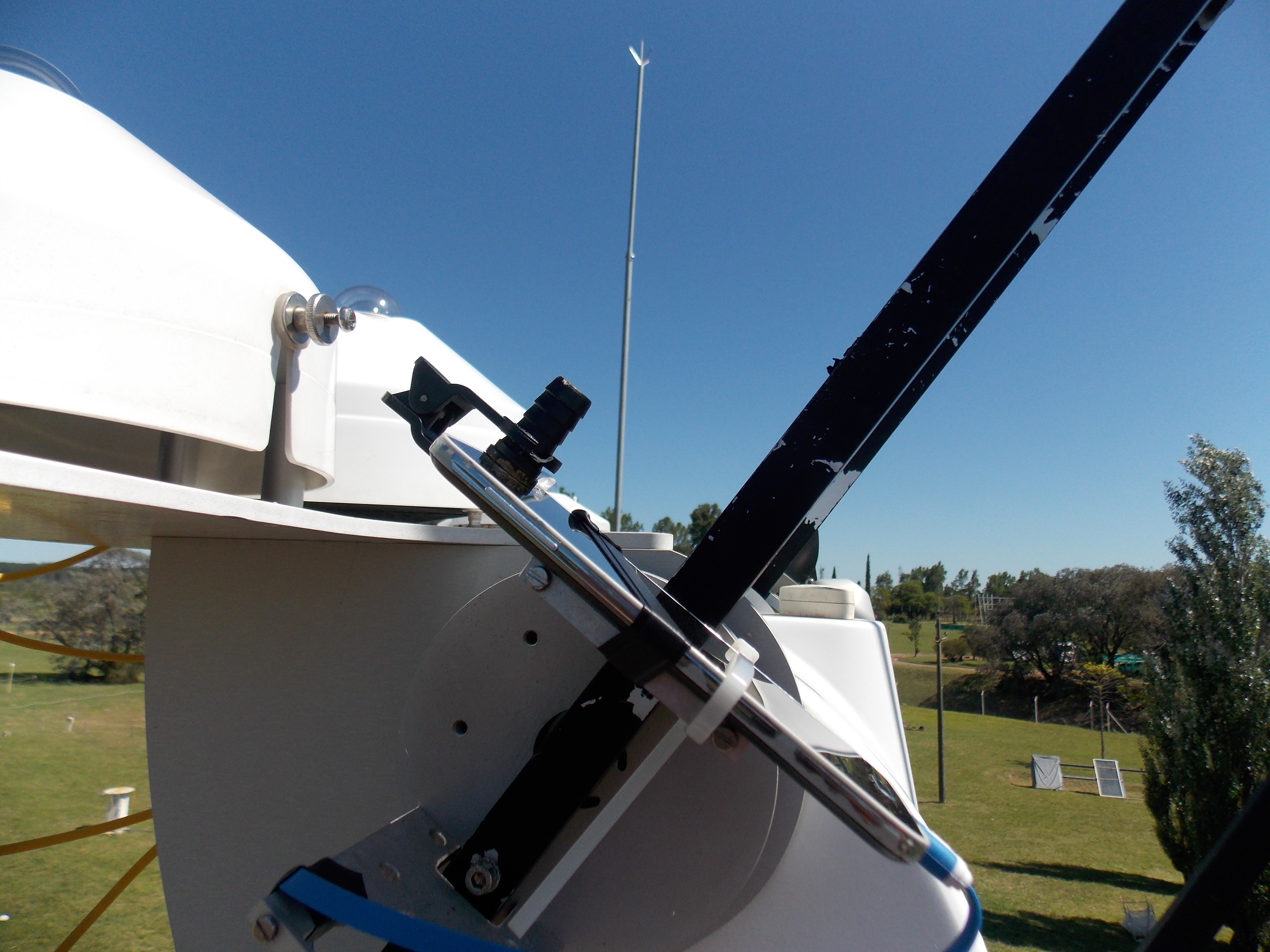}
\caption{Experimental setup. In the left panel the professional
  reference equipment (pyrheliometer) and the smartphone can be
  appreciated. The right panel offers a different perspective of the
  smartphone alignment.} \label{fig:medida_experimental}
\end{figure*}

The measurements were made at the %\textcolor{red}{(anon\-y\-mized).}
Solar Energy Laboratory (LES) of the University of the Republic
(Udelar). The experimental site of this laboratory is located in Salto
department, in northwestern Uruguay, with geographic coordinates of
$\phi=-31.28^\circ$ (latitude) and $\psi=-57.92^\circ$ (longitude),
corresponding to the UTC-3 time zone.

\begin{figure*}[ht!]
\centering
%\subfloat[Smartphone and diffusers.]
\includegraphics[height=0.56\textwidth]{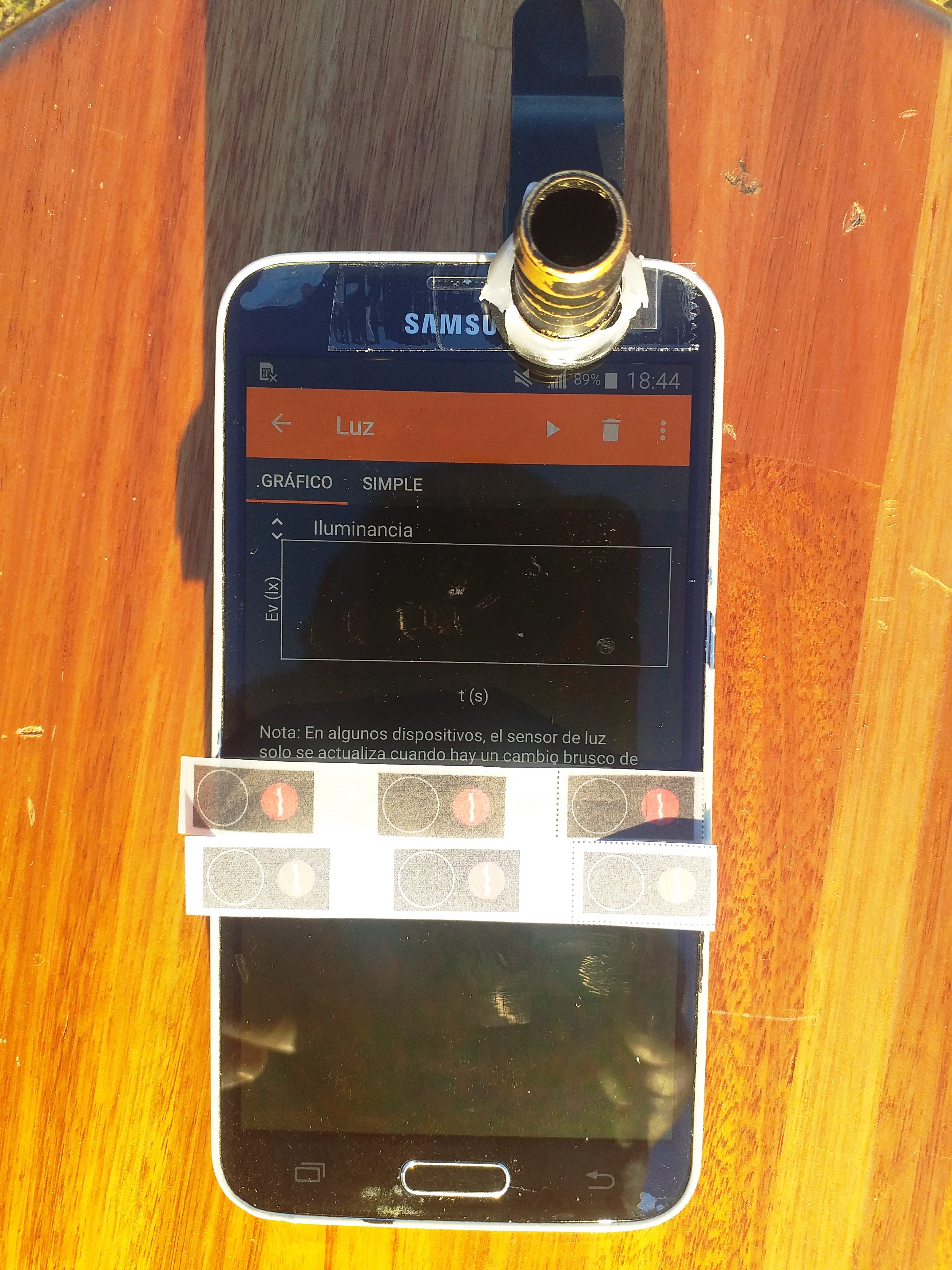}\hspace{0.5mm}
%\subfloat[Screen protection to prevent overheating.]
\includegraphics[height=0.56\textwidth]{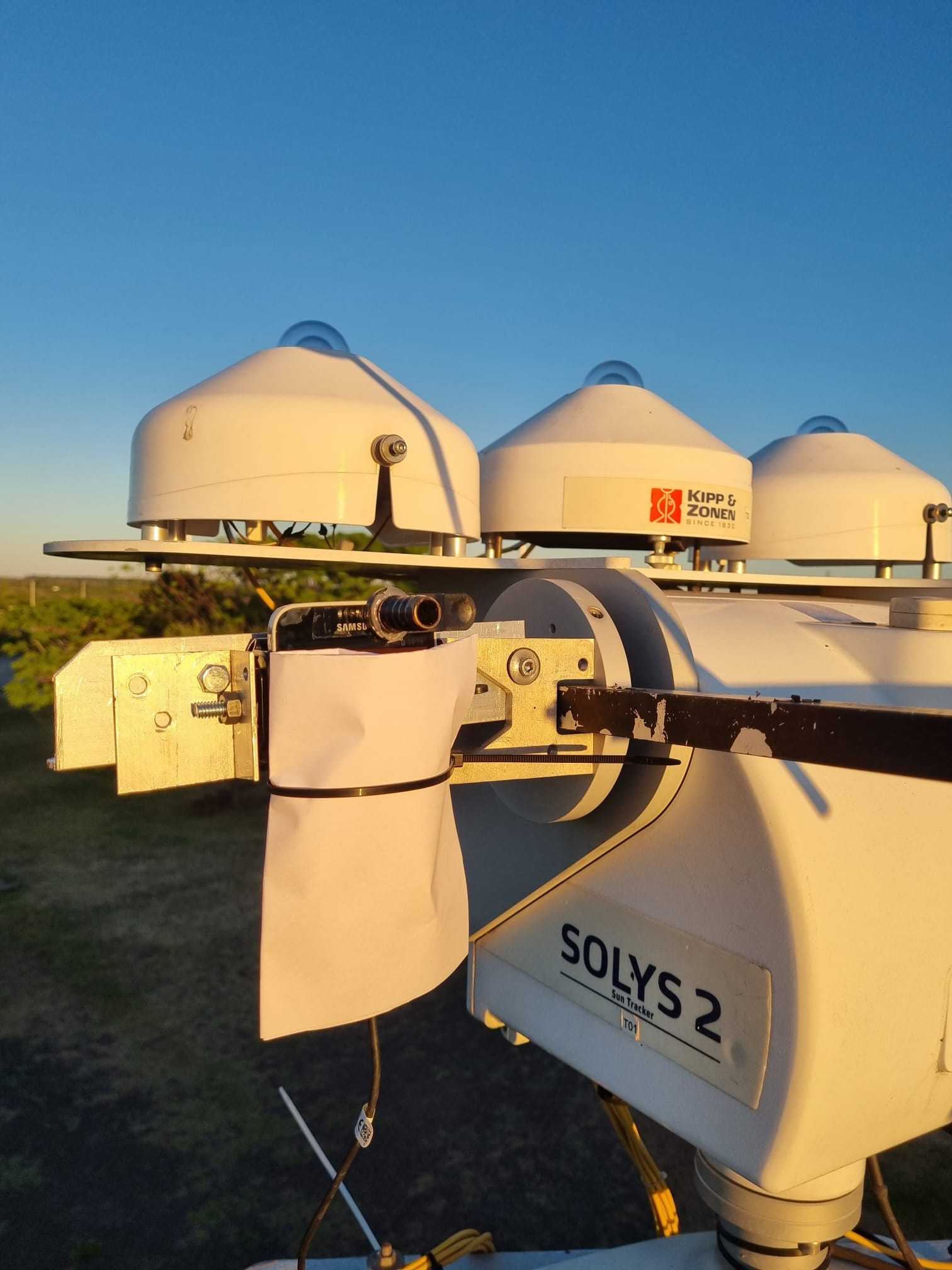}
\caption{Detailed view of the experimental setup. The left panel
  exhibits the smartphone and the diffusers while the right panel
  shows the screen protection used to prevent overheating.}
\label{fig:celular}
\end{figure*}

In this experiment, the ambient light sensor of smartphone Samsung S5
were used thanks to the freely available Phyphox app \cite{phyphox}.
A simple diffuser is placed above the sensor, in this case, tracing
paper printed in black, which prevents saturation of the recorded
signal. A cylindrical tube painted black is also placed, which acts as
a collimator for a large part of the diffuse irradiance, emulating the
professional pyrheliometer collimator (see Fig.~\ref{fig:celular}).
We emphasize that the selection of the diffuser offers no difficulties
since its only function is to attenuate the radiation to avoid
saturation and therefore does not affect the calibration procedure.
Lighting measurements with the light sensor are recorded by the
Phyphox app on a minute scale. Once the smartphone is placed in the
solar tracker, the collection begins keeping the device measuring
during a interval from several minutes to hours. To protect the phone
screen while the measurements are recorded throughout the day, a
double sheet of white paper (A4) placed in front of the smartphone
screen, acting as a radiation blocker to prevent the device from
overheating as shown in Fig.~\ref{fig:celular}.

The selection of an appropriate diffuser is important in achieving
accurate measurements with a smartphone. The smartphone's
analog-digital converter incorporates internal electronics that adjust
its gain based on the illuminance detected by the
sensor. Consequently, if the solar radiation measurements are low
(below $10$ klx, in this case), the equipment will automatically
change its scale without notifying the user. Each scale is associated
with a specific sensor saturation value, and this scaling behavior can
result in erroneous measurements for significant periods when the
measurement is in close proximity to the saturation
value. Furthermore, for the specific smartphone used in this study,
the upper limit of measurable illuminance in the higher range scale is
$60$ klx, which represents the saturation threshold. As a point of
reference, Michael et al. \cite{constante} obtained a conversion
constant of $120$ lx / W m$^{-2}$, indicating that measuring 1000
W/m$^{2}$ would not be feasible with our smartphone (approximately
$120$ klx) without the inclusion of a diffuser. Hence, it is essential
to regulate the attenuation of illuminance before it reaches the
sensor, for two primary reasons: (i) to ensure that values can be
accurately recorded without saturating the sensor, and (ii) to
maintain consistent measurements within a specific range of scales at
all times. This implies that the introduced attenuation by the
diffuser must strike a balance, neither being too minimal nor too
excessive, but rather falling within an intermediate range.

\begin{figure*}[ht!]
\centering \includegraphics[width=0.49\textwidth]{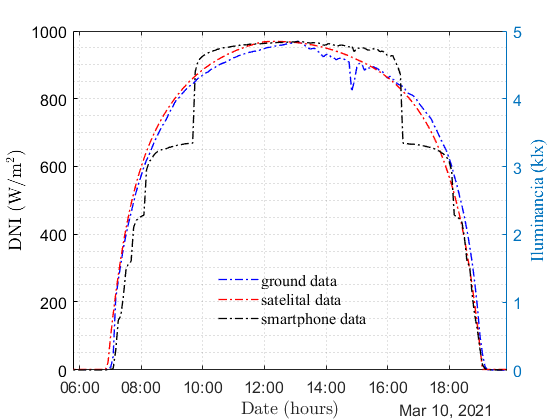}
\includegraphics[width=0.49\textwidth]{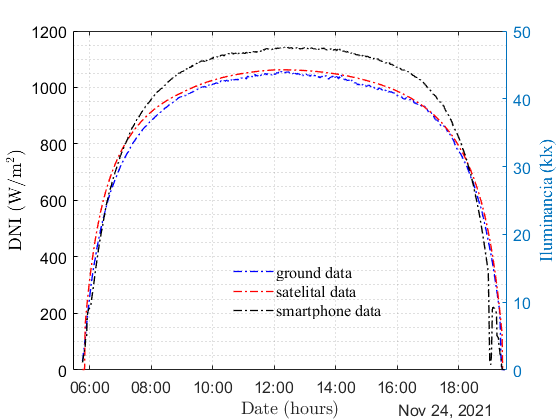}
\caption{Daily temporal evolution of the direct normal irradiance
  (DNI) using the smartphone with different diffusers: ordinary white
  paper (left) and black printed tracing paper (right).}
\label{fig:difusores}
\end{figure*}

To illustrate the impact of different diffusers, we present the
results obtained from two clear-sky days using two distinct types of
diffusers, as depicted in Fig.~\ref{fig:difusores}. The diffusers
employed were: (a) ordinary white paper with a surface mass density of
120 g/m$^2$, and (b) black printed tracing paper. The graph indicates
the illuminance measurements captured by the smartphone (indicated in
black) with each diffuser, alongside the reference direct irradiance
measurements obtained from the pyrheliometer (displayed in blue), and
the clear-sky satellite estimates (depicted in red). The behavior of
the measurement obtained using diffuser (a) is illustrated in the left
panel of Fig.~\ref{fig:difusores}, where the various scale changes
occurring at low illuminance levels between $0$ and $5$ klx are
clearly observed, along with the corresponding saturation points on
each scale. A similar behavior at low illuminance can be observed for
diffuser (b) in the right panel of Fig.~\ref{fig:difusores}, but only
for values below $10$ klx, with notable prominence during sunset.  For
measurements within the range of values exceeding $10$ klx, the
equipment does not undergo scale changes, resulting in continuous and
seemingly anomaly-free measurements facilitated by diffuser (b).  It
is also evident from the graph that the ordinary white paper diffuser
attenuates the signal to a greater extent compared to the tracing
paper diffuser (as depicted on the right-hand side $y$-axis of both
plots). This difference can be attributed to the higher reflectivity
of ordinary white paper, particularly within the visible region of the
solar spectrum. Therefore, based on our findings, we recommend the use
of diffuser (b) in this study. Custom selections may be done for other
smartphones, however, this point requires special attention.

To validate this method, we compared it with reference equipment.
In this case,  the signal generated by the
CHP1 pyrheliometer (in mV) is recorded by a Fisher Scientific
DataTaker DT85 data logger and is converted to irradiance (in
W/m$^{2}$) through the equipment constant. This measurement is the
reference DNI measurement of the %\textcolor{red}{(anon\-y\-mized)}
LES lab, and it is recorded continuously on a minute scale as an
average of instantaneous measurements every 15 s.

The precise measurement of DNI presents some difficulties. To carry it
out, a pyrheliometer is used, an instrument that consists of an array
of thermocouples (pyranometer) attached to a collimator tube, and a
precision solar tracking mechanism. If the equipment is aligned with a
precision of less than $0.1$°, the pyrheliometer is capable of
measuring the DNI with an uncertainty about 1$\%$. The measurements
are carried out in broadband, that is, the irradiance corresponding to
wavelengths between 200-4000~nm (which includes the entire solar
spectrum) is integrated into a single value. Figure
\ref{fig:medida_experimental} shows the experimental setup of this
work consisting of a Kipp \& Zonen \cite{kippzone} CHP1 pyrheliometer
(blue oval) and a Samsung S5 smartphone (yellow oval) assembled in a
Solys2 precision solar tracker. The assembly of the smartphone is
shown in Fig.~\ref{fig:celular}, in particular, its location
perpendicular to the axis of the black bars and the assembly of the
small hand-made collimator tube for the light sensor.

\subsection{Calibration based on high precision clear-sky estimates}
As an alternative calibration method for places where a professional
DNI measurement is not available, it is possible to use as reference
accurate clear-sky estimates that use information from weather
satellites and physically-consistent atmospheric models. This change
in the reference implies a slight increase in the uncertainty in the
determination of the calibration constants, since the DNI satellite
estimate presents greater uncertainty than a ground reference
measurement. There are sophisticated clear-sky models that integrate
estimates of different atmospheric variables, either by satellite or
by atmospheric reanalysis models, which can be considered as reference
\cite{Gueymard-2008,Lefevre-2013} if they have been validated by
terrestrial measurements with good concordance in various parts of the
world.

One interesting choice is the CAMS \cite{Copernico} (Copernicus
Atmosphere Monitoring Service) platform which provides free clear-sky
estimates using one of these reference models for the entire globe,
the McClear model \cite{Lefevre-2013}. This model is based on
sophisticated radiative transfer calculations from the libRadTran
\cite{Mayer-2005} library and its operational version takes the form
of a multiple input table based on real-time information on the state
of the atmosphere. In particular, this model uses information on
aerosols, precipitable water column, and ozone obtained from the CAMS
reanalysis database and Earth albedo estimates obtained by the MODIS
low-orbit satellite. The CAMS reanalysis in turn assimilates weather
satellite's information to provide its modeled data. Its platform
enables access to 1-minute (and other time scales) solar irradiance
estimates (global, direct and diffuse) from this high-precision model
by simply entering the latitude and longitude of the site of
interest. The file header contains information on each of the solar
magnitudes provided. For example, the dimensions of radiation are
Wh~m$^{-2}$ (irradiation, energy in the time interval per unit area),
which must be converted to W~m$^{-2}$ (average power per unit area) by
the corresponding conversion depending on the time scale.

\section{Practical use of the smartphone light sensor}
\label{calibracion}
\begin{figure*}[ht!]
\centering
{\includegraphics[width=0.98\textwidth]{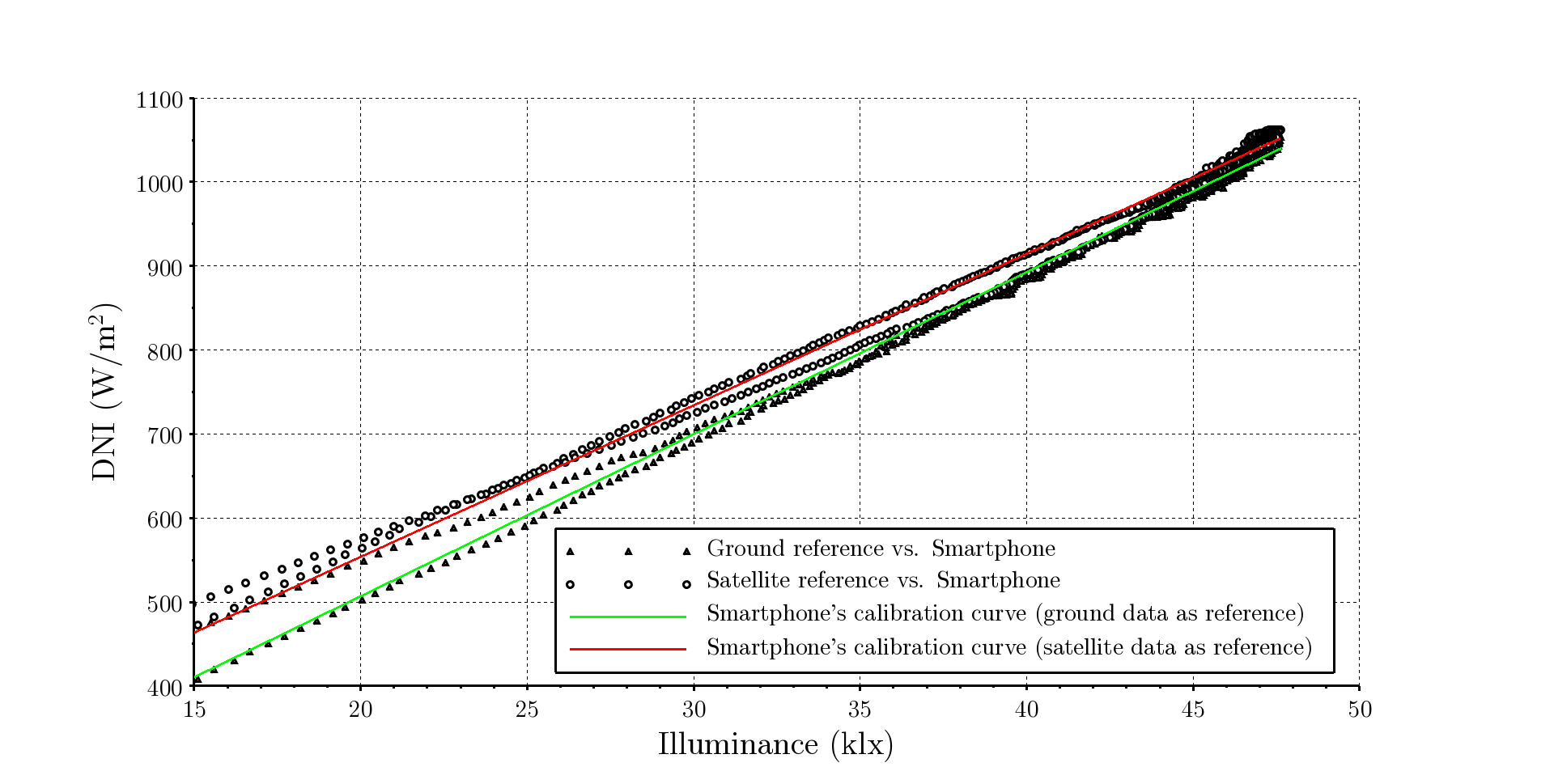}\label{fig:cal2}}
\caption{Calibration curves obtained from linear regression between
  smartphone measurements and reference DNI from terrestrial
  measurements and from McClear estimates.}
\label{fig:cal3}
\end{figure*}

After selecting a suitable diffuser (for example the black-printed
tracing paper used in this work) we can compare the smartphone
illuminance measurements with the DNI data. Here, a calibration
function
\begin{equation}
G_{bn} = a  E_{v,bn} + b
\label{eq:cal}
\end{equation}
will be used, where $E_{v,bn}$ is the illuminance measured by the
smartphone expressed in klx, $G_{bn}$ are the DNI data expressed in
W~m$^{-2}$, and $a$ and $ b$ are two conversion constants to adjust.
The calibration is performed with the two reference DNI data sets
considered, the professional measurement of the pyrheliometer and the
estimates of the McClear model.  Minute measurements and estimates of
November 24, 2021 were used at the %\textcolor{red}{(anon\-y\-mized)}
LES experimental site, where clear-sky conditions were maintained
throughout the day. Observing the plot of panel b in
Fig.~\ref{fig:difusores}, for the adjustment we used only the data
that meet $E_{v,bn} > 15$~klx, so that the smartphone sensor was
always on the same scale of measurement. This value was chosen
conservatively, in order to ensure measurements at intermediate values
on the scale. The calibration constants for both cases are presented
in Table~\ref{tab:cal3} and the experimental fit in
Fig.~\ref{fig:cal3}.

\begin{table}[ht!]
\centering
\setlength{\tabcolsep}{12pt}
\begin{tabular}{c|c|c|}\hline
\multicolumn{1}{|c|}{\textbf{Calibration}}& \textbf{Ground }&
\textbf{McClear } \\ \multicolumn{1}{|c|}{\textbf{constant}}&
\textbf{measurement}& \textbf{estimation}\\ \hline
\multicolumn{1}{|c|}{$a$ (W~m$^{-2}$ / klx)} &19.29 & 18.04
\\ \multicolumn{1}{|c|}{$b$ (W~m$^{-2}$)} &121.1 & 193.0\\ \hline
\multicolumn{1}{|c|}{Uncertainty in $a$ (W~m$^{-2}$ / klx)} &0.05
&0.05 \\ \multicolumn{1}{|c|}{Uncertainty in $b$ (W~m$^{-2}$)} &2.2 &
2.5\\ \hline \multicolumn{1}{|c|}{Relative uncertainty ($2\sigma$
  interval) in $a$} &0.54\% &0.65\%\\ \multicolumn{1}{|c|}{Relative
  uncertainty ($2\sigma$ interval) in $b$} &3.7\% & 2.6\%\\ \hline
\end{tabular}
\caption{Calibration constants from  Eq. \ref{eq:cal} corresponding to pyroheliometer data or satellite models
takn as references.}
\label{tab:cal3}
\end{table}

Figure~\ref{tab:cal3} reveals that the constants $a$ and $b$ can be
determined with low statistical uncertainty for each reference data
set (terrestrial measurements and McClear estimates). These
uncertainties have been obtained from the linear regression assuming a
Gaussian distribution of fluctuations. The table presents the
statistical uncertainties in each parameter, both absolute and
relative respect to its value, the latter for an interval of
$2\sigma$, which approximately represents a confidence level of
95\%. For this confidence level, the uncertainty in $a$ is less than
1\% and of $b$ less than 4\%, for both data sets.

\begin{figure*}[ht!]
\centering
\includegraphics[width=0.99\textwidth]{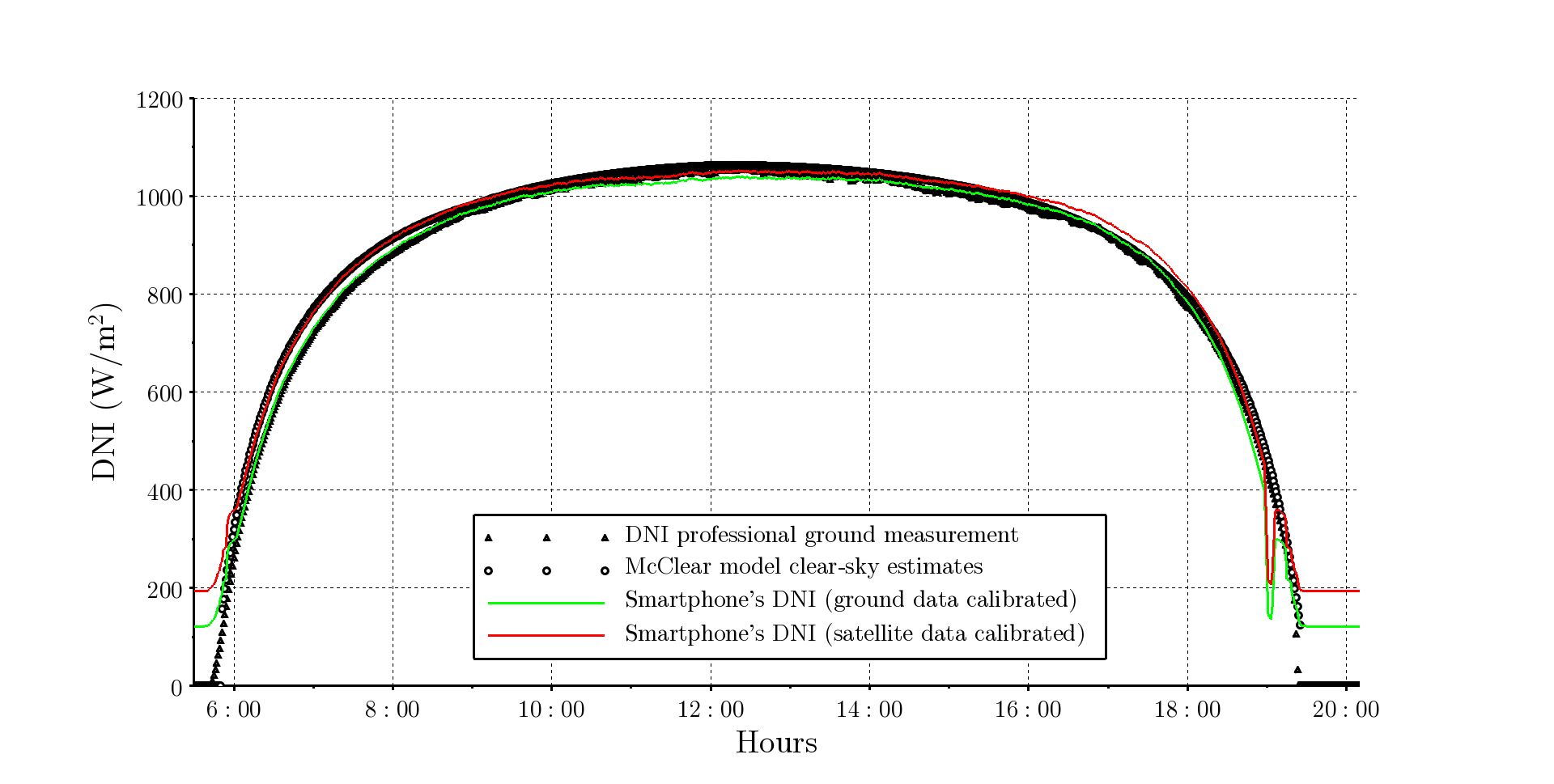}\label{fig:smart_sat}
\caption{DNI measurements obtained with smartphone on November 24, 2021.}
\label{fig:smart}
\end{figure*}

It is interesting to compare the calibration curves. Satellite
estimates exhibit deviations from the terrestrial measurements, so the
calibration curve based on these data will present more uncertainty
than the one obtained by comparison with ground measurements. As
observed in Fig. \ref{fig:difusores}, the McClear estimates for that
day overestimate the direct irradiance. This leads the calibration
curve obtained with this data set results above the calibration curve
obtained with terrestrial measurements as observed in
Fig.~\ref{fig:cal3}. Comparison of the McClear calibration curve with
the measured DNI data reports a mean deviation of +2.1\%
(overestimation) and a mean square deviation of 2.6\%. This
uncertainty is above the measurement uncertainty of the reference
instrument (1\%), so it is distinguishable, but at the same time it is
below the typical uncertainty of clear-sky satellite models (3-6\%)
\cite{Laguarda-2020}. This demonstrates that it is possible to perform
the calibration based on satellite data of solar irradiance with a low
uncertainty, enabling its use in the absence of reference
measurements. Figure \ref{fig:smart} shows the measurements obtained
with the smartphone using both calibrations. As can be seen, in both
cases a very good DNI measurement is achieved and completely
acceptable for an instructional laboratory for a wide range of the
day. The only downside is that the smartphone does not achieve a good
measurement in the first and last minutes of the day where the
illuminance is very low. Outside the diurnal range, i.e. when the
sensor measurement is zero, the DNI measurement is affected by the
non-zero offset of Eq.~\ref{eq:cal}.

These experimental results reveal that the smartphone is an
outstanding tool for measuring solar irradiance through its
illuminance sensor, applicable in low-cost university physics
laboratories. Furthermore, the measurement capacity a\-chieved with
the equipment is really good, even in comparison to commercial
sensors. Two relevant questions regarding the measurement capacity of
the smartphone arises here. The first is to evaluate the typical
uncertainty of a smartphone sensor used to measure DNI with both
calibration methods. To answer this question would require data
acquisition for several consecutive days (2-3 weeks) similar to
professional calibrations following international standards
\cite{ISO9847}. The second is to evaluate the stability of the
calibration curve over time, that is, how robust is the sensor to
gradual degradation. This would require a professional calibration of
the sensor every 3 months for about a year. With this set of tests it
is possible to technically evaluate the capacity limits of the
smartphone sensor for the solar irradiance measurement. In fact,
moderate-cost professional irradiance sensors are recommended to be
calibrated once a year, and a similar recommendation can arise for the
smartphone sensor. We recommend here to perform the smartphone
calibration each time before its utilization (i.e. the first day of
measurements). Similar studies can be carried out for the measurement
of global irradiance in the horizontal plane, not only direct
irradiance, which surely requires evaluating the non-planar angular
response of the smartphone sensor.

\section{Conclusion}\label{conclusion}

This study reveals the remarkable potential of smartphone light
sensors as effective tools for measuring direct solar irradiance and
for introducing students to the fundamental aspects of the solar
resource.  To validate the possibility of calibrating smartphones,
this experiment was conducted using professional solar measurement and
tracking equipment t Once this feasibility was demonstrated,
instructors could guide students to prepare their own smartphones
similarly to the experiment, using tracing paper and a simple
collimator tube, to measure direct illuminance in normal
incidence. These measurements, along with direct solar irradiance data
measured on-site or from satellite estimates, could be used to
accurately calibrate the smartphones, providing an innovative tool to
advance understanding of solar resources.  Remarkably, this work
demonstrated that calibration against both data sets can be performed
without introducing a significant increase in uncertainty, resulting
in highly reliable measurement capabilities suitable for instructional
laboratories. It is important to conduct calibration on a clear-sky
day, following the same principles applied in professional
calibrations according to current ISO standards for commercial
radiometers.

Our findings pave the way for the development of various
low-cost instructional laboratories, both within traditional classroom
settings and in outdoor environments.  In the process, students must
learn to develop important scientific-mathematical skills such as the
recording and processing of experimental data, assess their quality
with or without data filtering, and assigning uncertainty to
simplified models for direct solar irradiance estimation at ground
level. By leveraging smartphone technology, students can gain practical insights
into solar irradiance measurement, fostering a deeper understanding of
this important aspect of renewable energy resources.

%%%%%%%%%%%%%%%%%%%%%
\authorcontributions{Conceptualization, J.L.D.L. and R.A.S.; methodology, J.L.D.L. and R.A.S.; software, A.M. ; validation, J.L.D.L. and R.A.S and M.M.; data curation, J.L.D.L. and M.M.; writing---original draft preparation, J.L.D.L.; R.A.S and A.C.M.; writing---review and editing, J.L.D.L.; R.A.S and A.C.M.; visualization, J.L.D.L. and M.M.  All authors have read and agreed to the published version of the manuscript.}

\funding{This research received no external funding.}

\institutionalreview{Not applicable.}

\informedconsent{Not applicable.}

\dataavailability{ The data that support the findings of this study
  are available on reasonable request.}

\acknowledgments{We thank Sebastian Staacks, developer of PhyPhox, who
  especially made a more accurate version of the app.}

\conflictsofinterest{The authors declare no conflicts of interest.}

%%%%%%%%%%%%%%%%%%%%%%%%%%%%%%%%%%%%%%%%%%
%% Optional

%% Only for journal Encyclopedia
%\entrylink{The Link to this entry published on the encyclopedia platform.}

\abbreviations{Abbreviations}{
The following abbreviations are used in this manuscript:\\

\noindent 
\begin{tabular}{@{}ll}
MDPI & Multidisciplinary Digital Publishing Institute\\
DOAJ & Directory of open access journals\\
TLA & Three letter acronym\\
LD & Linear dichroism
\end{tabular}
}

\reftitle{References}
\bibliography{solarinside2}

\end{document}